\documentclass[10pt,twocolumn,letterpaper]{article}

\usepackage{iccv}
\usepackage{times}
\usepackage{epsfig}
\usepackage{graphicx}
\usepackage{amsmath}
\usepackage{amssymb}
\usepackage{algorithm}
\usepackage{algorithmic}

\usepackage[breaklinks=true,bookmarks=false]{hyperref}

\iccvfinalcopy 

\def\iccvPaperID{****} 

\ificcvfinal\fi

\begin{document}

\title{Multi-Stage Pathological Image Classification using Semantic Segmentation}

\author{Shusuke Takahama$^1$~~~~Yusuke Kurose$^{1,2}$~~~~Yusuke Mukuta$^{1,2}$~~~~Hiroyuki Abe$^{1,3}$\\Masashi Fukayama$^{1,3}$~~~Akihiko Yoshizawa$^{3,4}$~~~Masanobu Kitagawa$^{3,5}$~~~Tatsuya Harada$^{1,2,6}$
\vspace{2mm}
\\$^1$~The University of Tokyo~~~$^2$~RIKEN~~~$^3$~The Japanese Society of Pathology\\$^4$~Kyoto University~~~$^5$~Tokyo Medical and Dental University\\$^6$~Research Center for Medical Bigdata, National Institute of Informatics\\
{\tt\small $\{$takahama, kurose, mukuta, harada$\}$@mi.t.u-tokyo.ac.jp}
}

\maketitle
\ificcvfinal\thispagestyle{empty}\fi

\begin{abstract}
Histopathological image analysis is an essential process for the discovery of diseases such as cancer. However, it is challenging to train CNN on whole slide images (WSIs) of gigapixel resolution considering the available memory capacity. Most of the previous works divide high resolution WSIs into small image patches and separately input them into the model to classify it as a tumor or a normal tissue. However, patch-based classification uses only patch-scale local information but ignores the relationship between neighboring patches. If we consider the relationship of neighboring patches and global features, we can improve the classification performance. In this paper, we propose a new model structure combining the patch-based classification model and whole slide-scale segmentation model in order to improve the prediction performance of automatic pathological diagnosis. We extract patch features from the classification model and input them into the segmentation model to obtain a whole slide tumor probability heatmap. The classification model considers patch-scale local features, and the segmentation model can take global information into account. We also propose a new optimization method that retains gradient information and trains the model partially for end-to-end learning with limited GPU memory capacity. We apply our method to the tumor/normal prediction on WSIs and the classification performance is improved compared with the conventional patch-based method.
\end{abstract}

\section{Introduction}

\begin{figure}[t]
\begin{center}
\includegraphics[width=1.0\linewidth]{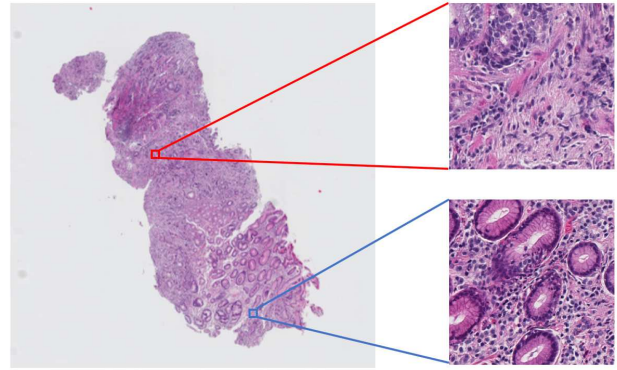}
\end{center}
\caption{Example of gigapixel WSI in the Stomach biopsy dataset. Pathologists need to scrutinize the details of WSIs, such as cellular tissue, for diagnosis. Therefore, WSIs must be of high resolution and large dimensions. When image analysis is performed on a WSI, small images called patches~(right) are cut out. Note that the tissue size, color, and shape exhibit substantial diversity depending on the location on the WSI.}
\label{fig:initial}
\end{figure}

Pathological diagnosis is the observation of tissue slides with a microscope to identify the presence of disease such as cancer. In recent years, a technology called digital pathology has been developed; it captures entire slide images with a scanner and stores it as a digital image~(whole slide image, WSI). Numerous WSI data has been accumulated with the spread of digital pathology, and many researches are aimed at assisting pathologists' diagnosis by applying image analysis technology using machine learning to WSI~\cite{survey1,path_first, path_patchbased, path_google, path_largescale1, path_camelyon16}.

The most characteristic aspect of WSI is the large image size and its exceptionally high resolution. Its size can be of over $10^5\times10^5$ pixels. In order to diagnose the presence of disease, it is necessary to investigate the local structures observed in the high resolution image. It is almost unfeasible to input WSIs directly into the machine learning model, considering GPU memory consumption. Furthermore, if we decrease the resolution in order to input the WSI into the model, the local features are lost, resulting in performance degradation. Therefore, past researches have adopted a method that divides a large WSI into small patch images and trains the classification model with the patches. In recent years, CNN has made significant achievements in the field of image recognition, and many researches in pathological image analysis also use CNN as a patch classifier ~\cite{path_patchbased,path_google}.

However, such patch-based method assesses each patch independently so that the relation between neighboring patches or more global information is not considered. The pathologist first inspects the whole WSI and zooms into the area suspected to be diseased, to scrutinize the details. Therefore, the classification performance is likely to be improved by considering both local and global features. 

Semantic segmentation is to perform local~(such as pixel) classification using features of broader area. Typical networks of semantic segmentation have a structure for estimating the class of each pixel while considering global features~\cite{fcn, U-net}. The performance of pathological image analysis is likely to improve with the model that considers global features and local information. However, because WSIs are exceptionally large and of very high resolution, the available memory capacity of GPU is insufficient for inputting WSIs directly into a segmentation model.

In this paper, we propose a model combining a feature extractor model considering local features and a segmentation model considering global information, in order to improve the classification performance of pathological analysis while regulating memory consumption. First, we input patches cut out from WSIs into the feature extractor model and extract patch features. Next, we collect features from each WSI, input them into the segmentation model, and obtain a whole slide tumor probability heatmap. We also propose a method to retain gradient information and train the feature extractor model partially, in order to optimize our network end-to-end through limited memory consumption. Using the proposed learning method, we trained using WSIs at its original resolution with low memory consumption.

Our main contributions in this paper: (1) We proposed a pathological classification model considering both high resolution local information and whole slide-scale global information. (2) We proposed a method to train the proposed model using lower memory consumption by retaining features and gradients between the two models and training the classification model separately. (3) We experimentally demonstrated that we achieved a higher classification accuracy using our model compared with the conventional patch-based method.

\section{Related Work}
{\bf Pathological~Classification:}~In previous pathological image analysis, manually designed features are extracted to classify the tumor/normal tissue, such as fractal features~\cite{path_old_fractal}, morphometric features~\cite{path_old_morpho}, and textural features~\cite{path_old_texture}. However, owing to the recent breakthrough of CNN in image recognition, many researches use CNN as a classifier~\cite{path_first, path_patchbased, path_google, path_largescale1, path_camelyon16}. Most of the researches adopts a method of dividing exceptionally large WSIs into small patches and training classification models with patches. In particular, various methods using CNN were proposed in Camelyon Grand Challenge~\cite{Camelyon17plus}, a competition of lymph node metathesis classification. Lee et al.~\cite{path_camelyon17}, who achieved the highest score, also use a CNN as the classifier. 
And some other techniques have been applied to improve the performance of the patch classification. Fine tuning is effective when having limited number of training data~\cite{path_finetune, path_largescale1}. Data augmentation is also important to improve generalization performance~\cite{path_google, path_patchbased}.

However, patch-based method does not consider the relation between neighboring patches or more global features. In order to consider peripheral information, Lee et al.~\cite{path_camelyon17} propose to apply processing such as smoothing to the prediction map. It helps the prediction score of neighboring patches to be spatially continuous. Liu et al.~\cite{path_google} input patches of multiple resolution into the model, in order to take spacial information into account. Tokunaga et al.~\cite{path_newest} train multiple classifiers for patches of different resolution and aggregate the outputs changing the weight of each CNN adaptively. However, these methods only use information of limited area around the patch, rather than the whole slide-scale global information. Furthremore, processing such as simple smoothing on the prediction map are likely to result in the omission of small disease areas.

\vspace{3mm}
{\bf Semantic Segmentation:}~Semantic segmentation is the classification of the local region~({\it e.g.,} pixel) categories considering the global information of images. Previous works used random forest or conditional random field~(CRF) as a classifier~\cite{crf-old, crf-new}. However, recently, most researches have been using classifiers based on CNN.

An early segmentation model using CNN is fully convolutional network~(FCN)~\cite{fcn}. FCN outputs a planar score map from an upsampled layer, instead of the CNN's final fully connected layer. SegNet~\cite{segnet} and U-net~\cite{U-net} have an encoder/decoder structure to retain the position information more sensitively. SegNet retains position information by recording the pooling index of the encoder; meanwhile, U-Net connects the intermediate feature map of the encoder to the decoder by the skip connection. These structures enable to output a high resolution prediction map considering both global and local features. Another structure for retaining position information is dilated convolution~\cite{dilated}; it can keep local information rather than a pooling layer. PSPNet~\cite{PSP_net} proposes a pyramid pooling module combining features down-sampled at different scales; meanwhile, DeepLabv3+~\cite{deeplabv3p} achieved cutting-edge performance by incorporating dilated convolution into PSPNet and adding a decoder structure. 

Based on the success of the semantic segmentation model, researches apply it also to pathological image analysis~\cite {path_seg_survey, path_seg, path_panoptic, path_largescale1}. In MICCAI 2015 Gland Segmentation Challenge Contest~\cite{path_seg_survey}, participants competed based on the accuracy of the semantic segmentation of a gland in pathological images. Chen et al.~\cite{path_seg}, who won the challenge, proposed a multi-task segmentation model based on FCN. It detects the boundary of a gland as well as performs normal pixel segmentation. However, this task detects the local tissue structure inside the small patches. It is challenging to apply the normal segmentation strategy to a gigapixel WSI because of memory constraints. Xu et al.\cite{path_largescale1} performs segmentation of an entire slide image by cutting out patches with small strides and voting for each pixel depending on the prediction score of the surrounding patches. However, this is just an extension of patch classification to the neighboring area.
In this paper, we propose a classification model that considers both local and global information, with low memory consumption, by combining classification model and segmentation model.

\section{Method}

In this section, we propose a new model for pathological image classification. In Sec.~\ref{method-str}, we propose the structure of proposed model combining two models. In Sec.~\ref{method-opt}, we propose two optimization methods of the model, Separate Learning and End-to-End learning. Separate Learning is relatively easy to optimize and stable, while End-to-End learning fixes the weights of the model more optimally.

\subsection{Model~structure \label{method-str}}
In pathological image analysis, the prediction scores of each patch from the classifier are arranged to obtain a whole slide prediction map. Then, we consider each patch on the prediction map as something like a pixel of image. If we replace the pixels of images with the feature vectors of each patch, we can apply a segmentation model on the WSI in a similar manner as general segmentation task. With the patches as the minimum component, we can obtain the whole slide prediction score map. 

\begin{figure}[t]
\begin{center}
\includegraphics[width=1.0\linewidth]{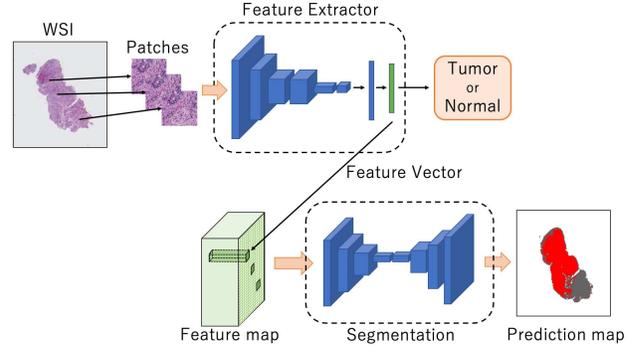}
\end{center}
\caption{Overview of proposed method. First, we extract the feature vector of the patches from the feature extractor model. Next, the feature vectors are arranged as a whole slide feature map. Finally, we input the feature maps into the segmentation model and obtain the tumor prediction maps.}
\label{fig:pipeline}
\end{figure}

Then, feature vectors representing patch information should be discriminative so that the model can classify tumor/normal. Therefore, we obtain discriminative feature vectors from the intermediate layer of the patch classifier. Specifically, first, we input patches from the WSI to the patch classifier and obtain the output of the intermediate layer as the feature vector of the patches. Next, we arrange the patch features based on their position on the WSI and develop a whole slide feature map. Finally, we input the feature map into the segmentation model and obtain a whole slide prediction map. The pipeline of the proposed model is shown in Figure~\ref{fig:pipeline}. Local information is considered in the patch classifier, and global feature is obtained in the segmentation model by integrating features with the encoder structure. The term ``global" in our paper refers to much broader area than patch-scale. Furthermore, we can reduce memory consumption by learning with the two-step model. 

We call the first half of the model as ``feature extractor model", which extracts features of the patches, the latter half as ``segmentation model".

\vspace {3mm}
{\bf Feature extractor model:}~Feature extractor model is applied to obtain the feature vector of patches. The input is the patches from the WSIs, which have tumor/normal labels based on the pathologists' annotation. The output of the final layer is the tumor/normal prediction score of the patches. When the layer of the model is defined as $f_1, f_2, ..., f_n$ in order from the input side, the output of $f_n$ is a two-dimensional~(tumor/normal) vector, and we extract the output of $f_{n-1}$ layer as the feature vector of the patches.

\vspace{3mm}
{\bf Feature~map~construction:}~The extracted feature vectors are arranged to obtain whole slide feature maps based on the position information where the patch is cut out. The feature map size is the number of cropped patches from the WSI. The depth of the map is the length of the feature vector. When we input the feature maps to the segmentation model, they should be of fixed size. However, the sizes of WSIs differ. Therefore, we first prepare a white map of sufficient size filled with zeroes. Then we arrange each feature maps so that they come to the center of white map.
We use this fixed feature map. Simultaneously, we create ground-truth label maps, in which the label information of the patches are arranged.


\begin{figure*}
\begin{center}
\includegraphics[width=1.0\linewidth]{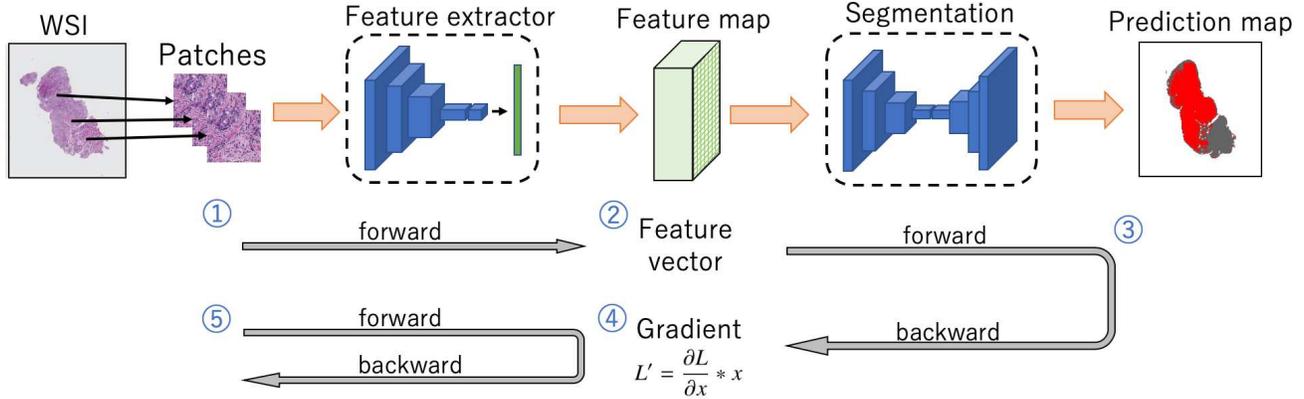}
\end{center}
\caption{Overview of End-to-End Learning method. To regulate memory consumption, we train the model part by part. First, the feature extractor models are calculated forward. The feature vectors of the last layer are retained and the outputs of the other layers are discarded. Next, the segmentation model is calculated forward and backproped using these vectors. The gradient information between the two models is held. Finally, the gradient information is used to train the feature extractor model. The model is calculated forward again and then backproped with the subset of patches.}
\label{fig:e2e}
\end{figure*}

\vspace{3mm}
{\bf Segmentation~model:}~The segmentation model integrates the features of patches with the encoder structure and performs classification considering global features. The input is the fixed size feature map, and the output is the tumor prediction score map of the WSI. The error function is the softmax cross entropy between the prediction map and ground-truth label map.

\subsection{optimization~method \label{method-opt}}
\subsubsection{Method1:~Separate~Learning}
As method 1, we propose Separate Learning. In this method, we train the feature extractor model and the segmentation model separately. We first optimize the feature extractor model with labeled patches from a WSI until it converges. Next, we fix the weight of the feature extractor model and input all the patches into the feature extractor model. With this process, the intermediate feature vector is extracted and the feature map of each WSI is developed. Then, the segmentation model is trained with the feature map. Finally, we obtain the whole slide prediction map as the output.

In this method, each model can be trained with limited memory consumption. We realize the segmentation of an exceptionally large WSI by learning separately with the two-step model.

\subsubsection{Method2:~End-to-End~Learning}

As method 2, we propose End-to-End Learning, which optimizes the feature extractor model and the segmentation model together. We estimate the memory consumption and explain how to train the model with exceptionally large WSIs as the input. When $N$ is the number of patches cut out from one WSI, $N \approx 10^3 \sim 10^4$. In order to input one feature map to the segmentation model, we must provide $N$ patches to the feature extractor model. When the memory consumption of the feature extractor model with one patch input is $M$, $M \approx 10^3$ MiB when using a image classification model such as VGG16~\cite{vgg}. If you attempt to input a WSI all at once, the feature extractor model consumes approximately $NM \approx 10^3$ GiB; this is unrealistic from the perspective of memory capacity.

There are several researches that split large CNN models and calculate partially, in order to regulate memory consumption~\cite{memory1, memory2, memory3, memory4}. Chen et al.\cite{memory2} divide a model into several blocks. Only the output of each block is retained and others are discarded, then the model again performs forward pass inside the block when obtaining the gradient. In our method, we consider the feature extractor model and segmentation model as two blocks. In the forward pass, only the feature vectors from the feature extractor model are retained and the intermediate layers' outputs are discarded. In the backprop pass, an additional forward pass is performed again. In addition, to further reduce memory consumption, $N$ patches constituting a WSI are divided into smaller subset of patches~(mini-batch) and calculated forward partially. If we divide $N$ patches into $r$ mini-batches and perform forward pass part by part, the memory consumption become approximately $\frac{NM}{r}$. By selecting an appropriate $r$ of computable size, we can calculate the feature vector with a low memory consumption. When all the feature vectors are obtained, we obtain an input to the segmentation model. For the backprop pass, the gradient from the segmentation model is used to calculate the gradient of the feature extractor model. At this time as well, the gradient is retained once between the two models. Moreover, the gradient of the feature extractor model is calculated for each mini-batch of size $r$; this reduces memory consumption. The gradient of the feature extractor model is calculated as follows:

In general, when the error $L$ is defined for the model, the updating formula of the model's weight $w$ is
\begin{eqnarray}
\label{eq:grad1}
w_{new} = w_{old} - \eta \frac{\partial L}{\partial w}
\end{eqnarray}
($\eta$ is a learning rate). If we can calculate $\frac{\partial L}{\partial w}$, we can update the weights. In the proposed model, the error $L$ is defined for the output of the segmentation model. Therefore, in order to calculate $\frac{\partial L}{\partial w_f}$~($w_f$ is the weight of the feature extractor model), we must retain the gradient between the two models.

We define the feature vector of the patches as $x$, which is the output of the feature extractor model. Then, $\frac{\partial L} {\partial w_f} $ is expressed using the chaining rule as
\begin{eqnarray}
\label{eq:grad2}
\frac{\partial L}{\partial w_f} = \frac{\partial L}{\partial x} \frac{\partial x}{\partial w_f}
\end{eqnarray}

Because $x$ is also the input of the segmentation model, we can calculate $\frac{\partial L}{\partial x}$. Using $\frac{\partial L}{\partial x}$, we define the new loss of the feature extractor model $L'$ as
\begin{eqnarray}
\label{eq:grad3}
L' = \frac{\partial L}{\partial x} \cdot x
\end{eqnarray}

We can calculate $\frac{\partial L}{\partial w_f}$ as
\begin{eqnarray}
\label{eq:grad4}
\frac{\partial L'}{\partial w_f} = \frac{\partial L}{\partial x} \frac{\partial x}{\partial w_f} = \frac{\partial L}{\partial w_f}
\end{eqnarray}

To summarize, we only have to calculate and retain $\frac{\partial L}{\partial x}$ of the segmentation model, and when optimizing the feature extractor model, calculate the inner product between the retained value and the output feature vector $x$. The value $L'$ can be considered as equivalent to the error of the feature extractor model and we can perform normal differentiation with $w_f$. Note that by defining the error $L'$ in this manner, the final layer $f_n$ of the feature extractor model is rendered unnecessary. Therefore, $f_n$ is removed in End-to-End Learning. 

The optimization procedure of End-to-End Learning is summarized as follows:

\begin{enumerate}
\renewcommand{\labelenumi}{\bf step \arabic{enumi}}
\item Calculate forward pass of feature extractor model and extract feature vector of patches. In this period, intermediate layers' outputs are discarded.
\item Arrange feature vectors to feature map considering their position information.
\item Perform forward and backprop pass of segmentation model with feature maps.
\item Calculate $L'=\frac{\partial L}{\partial x} \cdot x$ using segmentation loss $L$ and feature vector $x$.
\item Perform forward and backprop pass of feature extractor model and calculate gradient using $L'$.
\end{enumerate}

This procedure is shown in Figure~\ref{fig:e2e}. We repeat {\bf step 1} $\sim$ {\bf step 5} for each WSI. We reduce the memory consumption by discarding the intermediate layers' output and dividing the feature extractor model training. This method realizes segmentation at the scale of the high resolution WSI.

\section{Experiment}
In this section, we present the experimental result. In Sec.~\ref{ex-set}, we explain dataset details and experimantal settings. In Sec.~\ref{ex-res}, we present the results of three experiments. First, we evaluate the performance of the proposed method for patch-level classification on the Stomach biopsy dataset. Next, we similarly evaluate on the Camelyon16 dataset. Finally, we evaluate the patient-level classification performance using a Camelyon17 evaluation metric~\cite{Camelyon17plus}.

\subsection{Settings \label{ex-set}}
{\bf Dataset:}~We evaluate our proposed model on two types of datasets: Stomach biopsy dataset and Camelyon dataset. 
Stomach biopsy dataset that we prepared contains 1,019 WSIs of 996 patients. The maximum resolution is $\times$\nobreak20. The image size is approximately $10^4 \times 10^4$ pixels. The WSIs have tumor/normal pixel-level annotation for whole or a certain part of the tissue area. There are various reasons for the tissue area without annotation, such as the tissue deformation in the process of making slide or difficulty in determining the tumor/normal label. Dataset is randomly divided: 80$\%$ for training data and 20$\%$ for test data. Furthermore, the training data is divided: 80$\%$ for training data and 20$\%$ for validation data. 

Camelyon16 dataset~\cite{Camelyon16plus} and Camelyon17 dataset~\cite{Camelyon17plus} are lymph node tissue slides for detecting breast cancer metastases. The maximum resolution is $\times$\nobreak40. The image size is approximately $10^5 \times 10^5$ pixels. Camelyon16 dataset contains 400 WSIs, of these 270 are training data, and 130 are test data. Tumor/normal pixel-level annotation is provided for all the tissue areas of all the metastasis slides. The Camelyon17 dataset contains 1,000 WSIs of 200 patients: five slides per patient, and 500 slides for training data and 500 slides for test data. Only 50 WSIs in the training data has pixel-level region annotation. Each slide of the training data has slide level metastasis labels of four classes~(Macro, Micro, ITC, Negative) according to the size of the tumor area. Furthermore, based on the slide labels, the patient-level metastasis label~(pN-Stage) is defined. The assignment of the slide label and pN-stage follows the specified rule~\cite{Camelyon17plus}.

\begin{figure}[t]
\begin{center}
\includegraphics[width=1.0\linewidth]{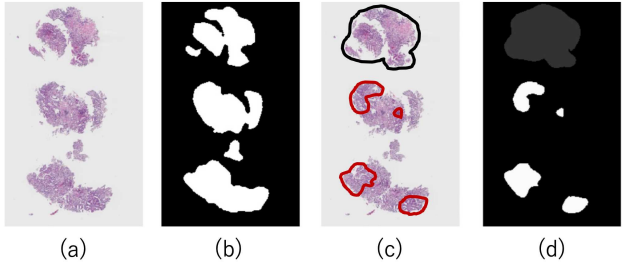}
\end{center}
   \caption{Process of patch extraction from WSI. (a) Original WSI (b) Tissue region extraction with Otsu threshold\cite{otsu} (c) Annotation by pathologist. The area surrounded by the black line is normal, and that surrounded by the line is a tumor. (d) Annotation mask generated from (c). White area is normal, gray area is tumor.}
\label{fig:slide}
\end{figure}

\vspace{3mm}
{\bf Pre-Processing:}~With regard to patch extraction from WSI, because WSI has a wide range of background~(no tissue) area, we separate the tissue area from the background area with Otsu threshold~\cite{otsu}. We cut out patches of $256 \times 256$ pixels without overlap and select those with over 80$\%$ pixels of tissue area. Next, based on the doctor's annotation, we label patches as ``tumor" if over 20$\%$ pixels is annotated as tumor. In non-tumor patches, we label as ``normal" if over 80$\%$ pixels are annotated normal. All the other patches are ``nolabel". The pre-processing is shown in Figure~\ref{fig:slide}.

From the Camelyon dataset, we can extract $10^4\sim 10^5$ patches from one WSI. Because there are much less tumor patches than normal ones~(approximately 2 $\%$ of normal in Camelyon16), we sampled an equal number of tumor/normal patches when training a feature extractor model, in order to deal with the imbalance between the classes. We could extract $10^3\sim 10^4$ patches from one WSI from the Stomach biopsy dataset. In this dataset, normal:tumor was approximately 4:1, therefore all the data was used without sampling.

Data augmentation is performed to improve the generalization performance. In training data, we add random cropping of the $224 \times 224$ pixel area, random rotation within an angle of [0, 90, 180, and 270], and random left--right flipping. In addition, to address the variation in the staining condition, color augmentation is added. Specifically, we change the saturation, contrast, brightness, and sharpness in the HSV space in the range [0.75, 1.25].

\vspace{3mm}
{\bf Implementation details:} We use GoogLeNet~\cite{googlenet} pretrained by Imagenet~\cite{imagenet} as the feature extractor model, and use U-Net~\cite{U-net} as the segmentation model. As the feature vector to extract, we add a fully connected layer before the final layer of GoogLeNet. The size of the feature vector is 16. The details of model structure analysis are written in supplementary material. In the optimization process of the segmentation model, we do not calculate the loss for the ``nolabel" patches. In the test process, the tumor/normal prediction is also assigned to the ``nolabel" patches. In End-to-End Learning, in order to stabilize learning, the feature extractor model and segmentation model use the weights learned in the process of Separate Learning. 

The feature map as an input of the segmentation model needs to be of a fixed size. We define a map size of $32 \times 32$ pixels for the Stomach biopsy dataset and of $512 \times 512$ pixels for the Camelyon dataset, considering the size of the tissue area on the slide. Because there are some lumps of tissue in the WSI of the Stomach biopsy dataset~(seen in Figure~\ref{fig:slide}), we develop a feature map for each lump of tissue. If the tissue size exceeds the fixed map size, the edge area of the tissue protrudes. In this experiment, we select a map size that fits almost all WSI tissue area, and it is also reasonable to increase the feature map size or resize the feature map of tissues to fit all tissue sizes.

We trained our networks with stochastic gradient descent in Chainer~\cite{chainer} and used the Adam optimization method~\cite{adam}.
In Separate Learning, the learning rate of the feature extractor model is 1e-4, and the model is trained 30 epochs with a batch size of 128. 
The learning rate of the segmentation model is 1e-4, and the model is trained 50 epochs with a batch size of 32.
In End-to-End Learning, the learning rate is 1e-9 for the feature extractor model and 1e-7 for the segmentation model. The model is trained 10 epochs.
We performed each method three times with different initial weights and evaluated with the mean value and the standard deviation.

\vspace{3mm}
{\bf Evaluation:}~In this method, we evaluate the performance of the tumor/normal binary patch classification. Specifically, we evaluate the classification accuracy and AUC of the precision recall curve~(PR-AUC). We set the tumor/normal threshold of accuracy evaluation as 0.5. When there is a large imbalance between classes, the PR-AUC is less susceptible to the bias and reflects the performance more precisely. ``Nolabel" patches are not used for evaluation. Furthermore, the classification performance is qualitatively evaluated by the prediction map outputted from the models.

On the other hand, Camelyon17 Grand Challenge evaluates the pN-Stage, which is a patient-level evaluation rather than a patch-level evaluation. First, the size of the tumor region of the prediction map determines the slide-level class prediction. Next, the pN-Stage is determined based on the slide-level prediction of five slides per patient. The prediction results are evaluated by kappa score~\cite{Camelyon17plus}. In the experiment on the Camelyon17 dataset, we compare the performance using this evaluation metric.

\subsection{Result \label{ex-res}}

\subsubsection{Evaluation on Stomach biopsy dataset \label{result-stomach}}

\begin{figure}[t]
\begin{center}
\includegraphics[width=1.0\linewidth]{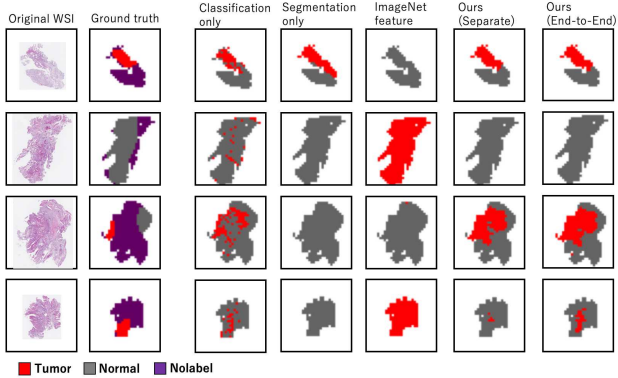}
\end{center}
\caption{Prediction score map output on Stomach biopsy dataset. Left to right: original WSI image, ground truth annotation map, Classification only, Segmentation only, Imagenet feature, Ours (Separate Learning), Ours (End-to-End Learning). Red area represents ``tumor", gray represents ``normal", purple is the ``nolabel" area, and white is the background area. The tumor/normal prediction is also assigned to the nolabel area. Proposed methods classify more precisely than others.}
\label{fig:result1}
\end{figure}

We compare the performance on the Stomach biopsy dataset. We evaluate the two proposed methods: Separate Learning and End-to-End Learning. We also evaluate three methods for comparison. (1){\bf Classification Only:} Similar to previous researches, we predict the patch score with only the classifier. We implement the classifier based on~\cite{path_camelyon17} method, which exhibited state-of-the-art performance in the Camelyon17 challenge, and evaluate it as a comparison method. (2){\bf Segmentation Only:} We classify the WSI score map with only the segmentation model. Because the original WSIs are excessively large to be inputted, we reduce the input tissue image to $1024\times 1024$ pixels to reduce memory consumption, which is $ \frac{1}{8}$ of the original resolution. We use U-Net as the segmentation model. (3){\bf Imagenet feature:} The feature extractor model is not trained by patches. We use the weight of GoogLeNet pretrained by Imagenet to extract the feature vectors and train the segmentation model. All the evaluations are performed based on accuracy and PR-AUC. The results of our experiment are presented in~\tablename~\ref{tb:result_comp1}. Furthermore, the prediction maps generated from each method are shown in Figure~\ref{fig:result1}.

The result demonstrates that the proposed methods outperform the comparison methods based on both accuracy and PR-AUC. In our method, the classification performance is improved by using both local and global information. Between the two proposed optimization methods, End-to-End Learning outperformed Separate Learning with respect to the PR-AUC. The performance appears to be improved by the optimization of the feature extractor model using the final loss of the segmentation model. We think the main reason for the slight difference between the performance of the two proposed methods is the high classification score of Separate learning, which leaves a limited room for improvement. 

\begin{table}
\begin{center}
\caption{Result on Stomach biopsy dataset}
  \begin{tabular}{c|c c} \hline
    Method & Accuracy($\%$) & PR-AUC($\%$)\\ \hline \hline
    Classifier Only & 95.92$\pm$0.11 & 96.08$\pm$0.10\\
    Segmenation Only & 92.24$\pm$0.14 & 87.21$\pm$1.85\\
    Imagenet feature & 89.35$\pm$0.59 & 78.99$\pm$1.59\\ \hline
    Separate Learning & {\bf 98.53$\pm$0.03} & 99.30$\pm$0.02\\
    End-to-End Learning & 98.45$\pm$0.01& {\bf 99.34$\pm$0.00}\\ \hline
  \end{tabular}
\label{tb:result_comp1}
\end{center}
\vspace{-6mm}
\end{table}

With regard to memory consumption, The training of the feature extractor model consumes approximately 8,000 MiB with a batch size of 128, and the training of the segmentation model consumes approximately 1,000 MiB with a batch size of 32. On the other hand, End-to-End Learning consumes only approximately 10,000 MiB. We make it feasible to train an entire WSI of high resolution by reducing memory consumption using our proposed method. Considering, the segmentation-only method, which reduces the resolution to $\frac{1}{8} $, consumes approximately 7,000 MiB with a batch size of 1, the proposed method realized significant reduction in memory consumption.

With regard to time consumption, We measured training time using NVIDIA DGX-1 (GPU: Tesla P100 16GB, CPU: Intel Xeon E5-2698 v4, RAM: 512GB). In classifier only method, model training takes about 1.0 hour per epoch. In Separate learning, feature map generation takes about 6.0 hours, and segmentation model training takes about 2.3 minutes for each epoch. End-to-end learning takes about 3.5 hours per epoch. Therefore, in the settings of our experiment, it takes 30 hours for classifier only, 38 hours for Separate learning, and 73 hours for End-to-end learning. Our method certainly takes more time compared with the other existing methods. However, even End-to-end learning model can be trained in about three days, and in medical field higher performance is more critical than training time.

Figure~\ref{fig:result1} shows that the classifier-only method captures an tumor part and achieves high accuracy. However, we observe that a ``tumor point", an independent tumor spot in a normal area, on the map. When supporting doctors' diagnosis in clinical situation, doctors have to scrutinize tissue that have even a slight tumor prediction, and such prediction map with many ``tumor point" cannot alleviate doctors' burden. Meanwhile, the proposed methods classify correctly and has no ``tumor point", that is a substantial advantage for doctors.

End-to-End Learning achieves highest performance but consumes more time for training and its training process is not stable because of its large model size. Therefore, there has to be a trade-off between the high accuracy of End-to-End Learning and the convenience of training of Separate Learning.

\subsubsection{Evaluation on Camelyon16 dataset \label{result-16}}

\begin{figure}[t]
\begin{center}
\includegraphics[width=1.0\linewidth]{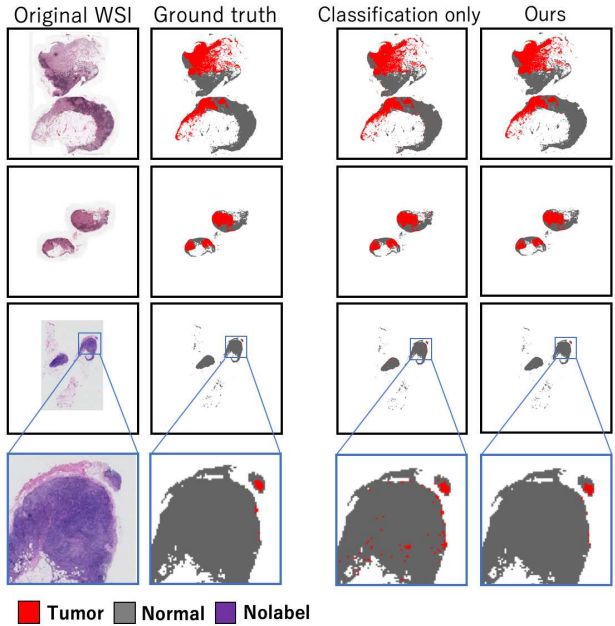}
\end{center}
   \caption{Prediction score map output on Camelyon16 dataset. Left to right: original WSI image, ground truth annotation map, classification only, Ours (End-to-end learning). The last row is an enlargement of the tissue area. Although the Camelyon16 dataset includes WSI having a very small tumor area, the proposed method identifies it correctly. The prediction map from the classification-only method exhibits a number of ``tumor point", whereas our method exhibits few.}
\label{fig:result2}
\end{figure}

We compare the performance on the Camelyon16 dataset, because all slide in Camelyon16 has tumor/normal pixel-level annotation. Considering the result on the Stomach biopsy dataset, we evaluate and compare the performance of the classifier-only method and our End-to-End Learning method. The training settings are fundamentally similar to those in the experiment on the Stomach biopsy dataset. However, the training data was sampled so that the number of tumor/normal patches would be similar and all the test data were used for the evaluation. The End-to-End Learning is performed 5 epochs. The result of the experiment is presented in~\tablename~\ref{tb:result_comp2}. The proposed method exhibits higher performance than the classifier-only method. The proposed method is apparently effective irrespective of the dataset. The generated prediction map is shown in Figure~\ref{fig:result2}.

\subsubsection{Evaluation on Camelyon17 dataset \label{result-17}}

We compare the performance on the Camelyon17 dataset. Because the label of the Camelyon17 testset is not open, we evaluate with the validation data, as in~\cite{path_camelyon17}. We set 43 patients' slides, which include the region-level annotation slide in the Camelyon17 training set, and all the Camelyon16 dataset as the training data. And we set the remaining 57 patients' slides in the Camelyon17 training set as the validation data. First, we train the classification model with the training data. Next, we predict the slide-level label by training random forest as used in~\cite{path_camelyon17}. Finally, we predict the patient-level pN-stage based on the Camelyon17 metric~\cite{Camelyon17plus} and evaluate with kappa score. We validate the performance with the validation data by 5-fold cross-validation setting. 

\tablename~\ref{tb:result_comp3} shows patient-level kappa score. Our proposed method also improves patient-level classification performance. Note that there is some difference between the score described in~\cite{path_camelyon17} and the comparison method we implemented. The score is expected to be improved with additional techniques used for the competition such as ensemble learning and parameter tuning in~\cite{path_camelyon17}.

\begin{table}
\begin{center}
\caption{Result on Camelyon16 dataset}
  \begin{tabular}{c|c c} \hline
    Method & Accuracy($\%$) & PR-AUC($\%$)\\ \hline \hline
    Classifier Only & 96.23$\pm$0.21 & 96.39$\pm$0.67\\
    Ours~(End-to-End) & {\bf 98.14$\pm$0.05} & {\bf 99.24$\pm$0.03}\\
    \hline
  \end{tabular}
\label{tb:result_comp2}
\end{center}
\vspace{-3mm}
\end{table}

\begin{table}
\begin{center}
\caption{Result on Camelyon17 dataset}
\small
  \begin{tabular}{c| c} \hline
    Method & Kappa score\\ \hline \hline
    Classifier only & 71.9\\
    Ours~(End-to-End) & {\bf 76.4}\\
    \hline
  \end{tabular}
\label{tb:result_comp3}
\end{center}
\vspace{-6mm}
\end{table}

\section{Conclusion}
We proposed a pathological image classification model combining feature extractor model and segmentation model. We also proposed the End-to-End Learning optimization method on high resolution pathological images. It consumes marginal memory by retaining the gradient and training the model with small a subset of patches. Using this model, we achieved higher performance on the tumor/normal classification of WSI, than other patch-based methods. Furthermore, the prediction maps generated by the model facilitate doctors in identifying and assessing tumor areas. In the future, we will improve the model structure to further enhance performance and accelerate learning. We will also apply it to other datasets.

\section*{Acknowledgements}
This work was partially supported by a Grant for ICT infrastructure establishment and implementation of artificial intelligence for clinical and medical research from the Japan Agency of Medical Research and Development AMED (JP18lk1010028) and JST CREST Grant Number JPMJCR1403. This work was performed as the work of the Japan Pathology AI Diagnostics Project (JP-AID) research group in the Japanese Society of Pathology.

{\small
\bibliographystyle{ieee_fullname}
\bibliography{iccvbib}
}

\end{document}